\documentclass{aa} 

\usepackage{graphicx}
\usepackage{color}
\usepackage{soul}
\usepackage{txfonts}
\usepackage{float}
\begin{document}

   \title{A concordant scenario to explain FU Ori from deep centimeter and millimeter interferometric observations}

   \subtitle{}

   \author{Hauyu Baobab Liu\inst{1}
          \and
           Eduard I. Vorobyov\inst{2}\fnmsep\inst{3} 
          \and 
          Ruobing Dong\inst{4}
          \and  
          Michael M. Dunham\inst{5}
          \and  
          Michihiro Takami\inst{6}
          \and  
          Roberto Galv\'an-Madrid\inst{7}
          \and  
          Jun Hashimoto\inst{8}
          \and  
          \'Agnes K\'osp\'al\inst{9}\fnmsep\inst{10}
          \and  
          Thomas Henning\inst{10}
          \and  
          Motohide Tamura\inst{8}\fnmsep\inst{11}\fnmsep\inst{12}
          \and  
          Luis F. Rodr\'iguez\inst{7}
          \and  
          Naomi Hirano\inst{6}
          \and  
          Yasuhiro Hasegawa\inst{13}
          \and  
          Misato Fukagawa\inst{14}
          \and  
          Carlos Carrasco-Gonzalez\inst{7}
          \and  
          Marco Tazzari\inst{15}
          }

   \institute{European Southern Observatory (ESO), Karl-Schwarzschild-Str. 2, D-85748 Garching, Germany \\
                 \email{baobabyoo@gmail.com}
          \and
          Department of Astrophysics, University of Vienna, Vienna 1180, Austria
          \and
          Research Institute of Physics, Southern Federal University, Rostov-on-Don 344090, Russia
          \and 
          Steward Observatory, University of Arizona, Tucson, AZ, 85721, USA
          \and 
          Department of Physics, State University of New York at Fredonia, 280 Central Ave, Fredonia, NY 14063
          \and
          Academia Sinica Institute of Astronomy and Astrophysics, P.O. Box 23-141, Taipei, 106 Taiwan 
          \and 
          Instituto de Radioastronom\'ia y Astrof\'isica, UNAM, A.P. 3-72, Xangari, Morelia, 58089, Mexico
          \and 
          Astrobiology Center, NINS, Tokyo 181-8588, Japan
          \and 
          Konkoly Observatory, Research Centre for Astronomy and Earth Sciences, Hungarian Academy of Sciences, P.O. Box 67, 1525 Budapest, Hungary
          \and 
          Max-Planck-Institut f\"or Astronomie, K\"onigstuhl 17, D-69117 Heidelberg, Germany
          \and 
          Department of Astronomy and RESCEU, University of Tokyo, Tokyo 113-8654, Japan
          \and 
          National Astronomical Observatory of Japan, NINS, Tokyo 181-8588, Japan
          \and 
          Jet Propulsion Laboratory, California Institute of Technology, Pasadena, CA 91109, USA 
          \and 
          Division of Particle and Astrophysical Science, Graduate School of Science, Nagoya University, Furo-cho, Chikusa-ku, Nagoya, Aichi 464-8602, Japan
          \and 
          Institute of Astronomy, Madingley Road, Cambridge, CB3 0HA, UK
             }

   \date{Received December XX, 2016; accepted January XX, 2017}

  \abstract
    {} 
   {The aim of this work is to constrain properties of the disk around the archetype FU Orionis object, FU Ori, with as good as $\sim$25 au resolution.}
   {We resolved FU Ori at 29-37 GHz using the Karl G. Jansky Very Largy Array (JVLA) in the A-array configuration, which provided the highest possible angular resolution to date at this frequency band ($\sim$0$\farcs$07). We also performed complementary JVLA 8-10 GHz observations, the Submillimeter Array (SMA) 224 GHz and 272 GHz observations, and compared with archival Atacama Large Millimeter Array (ALMA) 346 GHz observations to obtain the spectral energy distributions (SEDs).
   }
   {Our 8-10 GHz observations do not find evidence for the presence of thermal radio jets, and constrain the radio jet/wind flux to at least 90 times lower than the expected value from the previously reported bolometric luminosity-radio luminosity correlation. The emission at frequencies higher than 29 GHz may be dominated by the two spatially unresolved sources, which are located immediately around FU Ori and its companion FU Ori S, respectively. Their deconvolved radii at 33 GHz are only a few au, which is two orders of magnitude smaller in linear scale than the gaseous disk revealed by the previous Subaru-HiCIAO 1.6 $\mu$m coronagraphic polarization imaging observations. 
 We are struck by the fact that these two spatially compact sources contribute to over 50\% of the observed fluxes at 224 GHz, 272 GHz, and 346 GHz. 
The 8-346 GHz SEDs of FU Ori and FU Ori S cannot be fit by constant spectral indices (over frequency), although we cannot rule out that it is due to the time variability of their (sub)millimeter fluxes.
   } 
   {The more sophisticated models for SEDs considering the details of the observed spectral indices in the millimeter bands suggest that the $>$29 GHz emission is contributed by a combination of free-free emission from ionized gas, and thermal emission from optically thick and optically thin dust components. We hypothesize that dust in the innermost parts of the disks ($\lesssim$0.1 au) has been sublimated, and thus the disks are no more well shielded against the ionizing photons.
The estimated overall gas and dust mass based on SED modeling, can be as high as a fraction of a solar mass, which is adequate for developing disk gravitational instability.   
Our present explanation for the observational data is that the massive inflow of gas and dust due to disk gravitational instability or interaction with a companion/intruder, was piled up at the few au scale due to the development of a deadzone with negligible ionization. 
The piled up material subsequently triggered the thermal instability and the magnetorotational instability when the ionization fraction in the inner sub-au scale region exceeded a threshold value, leading to the high protostellar accretion rate.
}

   \keywords{Stars: formation --- radio continuum: ISM --- submillimeter: ISM --- stars: variables: T Tauri, Herbig Ae/Be
               }

\titlerunning{High angular resolution millimeter observations on FU Ori}

   \maketitle

\section{Introduction}

\begin{figure*}
\hspace{-0.4cm}
\begin{tabular}{p{9cm} p{9cm} }
\includegraphics[width=9cm]{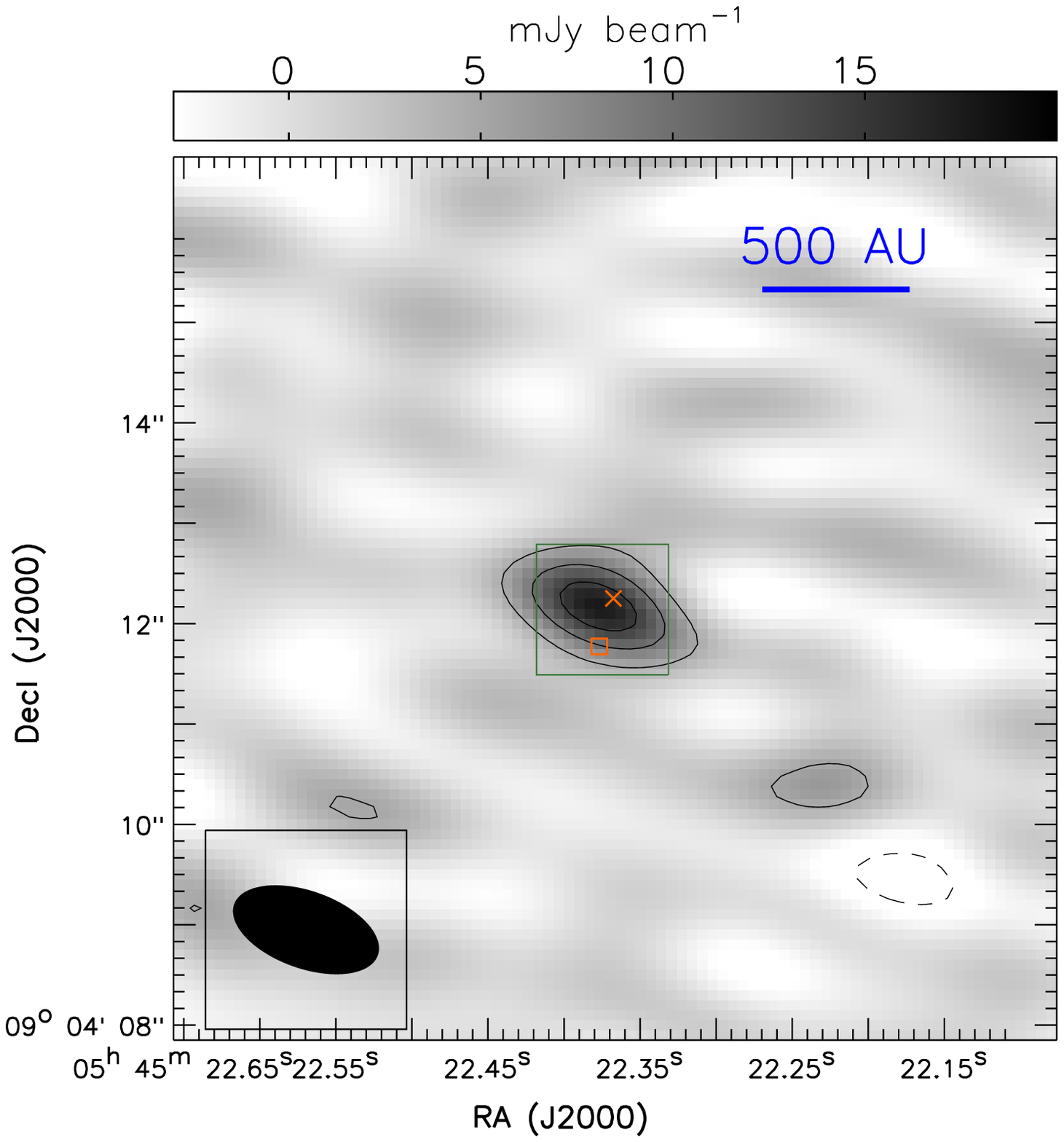} & \includegraphics[width=9cm]{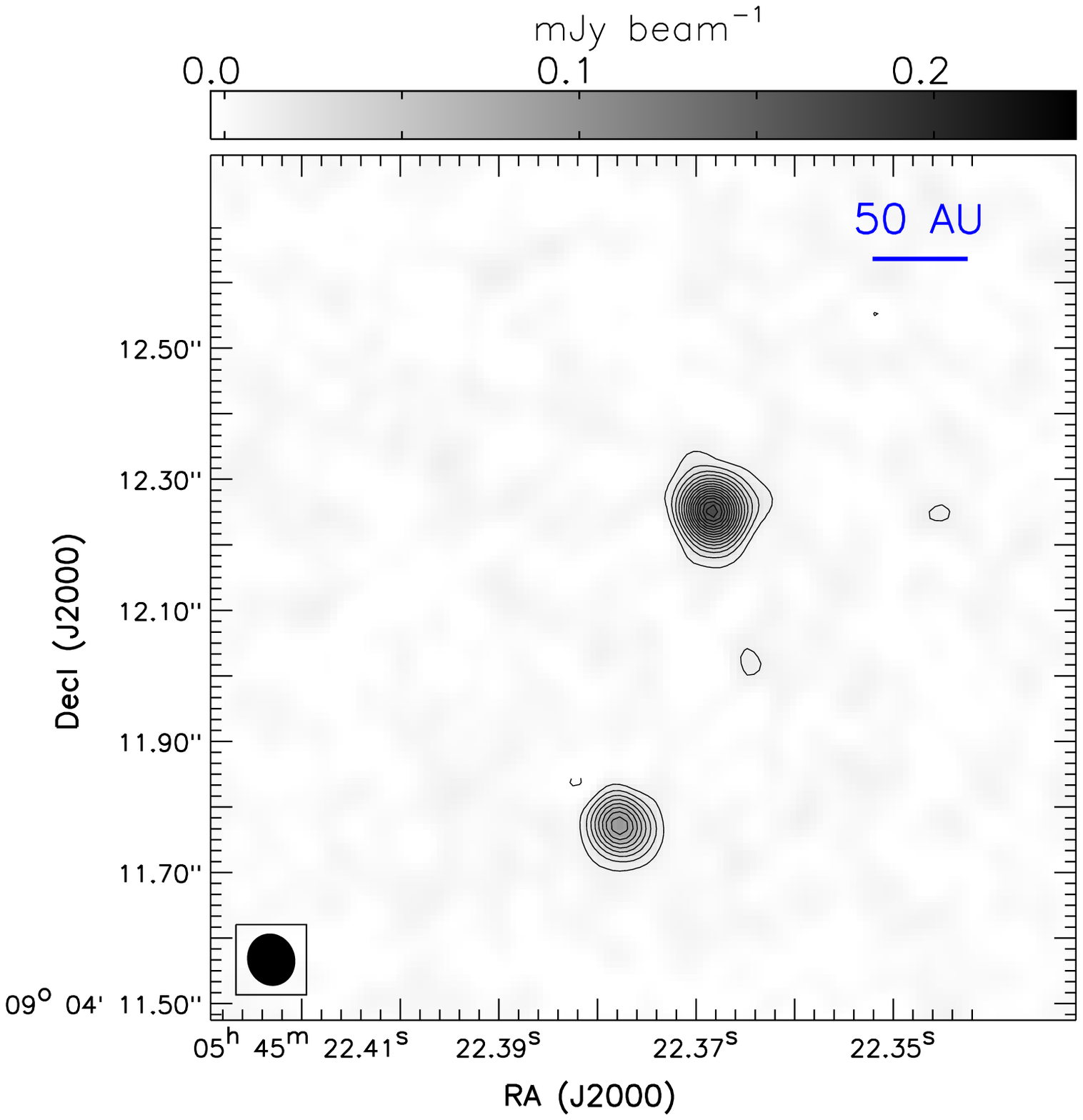} \\
\end{tabular}
\vspace{-0.1cm}
\caption{Interferometric images of FU Ori (contour and grayscale). 
Left:-- SMA 224 GHz image. Contour levels are 4.8 $m$Jy\,beam$^{-1}$ (3$\sigma$) $\times$ [-2, -1, 1, 2, 3]. 
Right:-- JVLA 29-37 GHz image. Contour levels are 10.8 $\mu$Jy\,beam$^{-1}$ (3$\sigma$) $\times$ [-2, -1, 1, 2, 3, 4, $\cdots$, 13, 14, 15].
Dashed contours present negative intensity.
Cross and square in the left panel mark the locations of FU Ori and FU Ori S, which were obtained from the 2-dimensional Gaussian fits to the JVLA 33 GHz image. 
Green box in the left panel shows the field of view of the right panel.
We assume a distance of 353 pc.
}
\label{fig:fuori}
\end{figure*}

The episodic accretion of young stellar objects (YSOs) may resolve the problem of the observed too low luminosities from Class 0/I YSOs as compared with what are expected from the stationary accretion models (Dunham \& Vorobyov 2012).
Indeed, optical and infrared monitoring observations have found accretion-outbursting YSOs (see Audard et al. 2014), some of which remain embedded within their circumstellar envelopes (Quanz et al. 2007).
The longer duration (few tens of years or longer) ones are named FU Orionis objects after the archetype source FU Ori (see Hartmann \& Kenyon 1996 for a review).
The shorter duration (few hundreds days to few years) and repetitive ones are named EXors after the archetype source EX Lupi (see Herbig 1977 for a review).
The physical mechanisms to trigger the accretion outbursts of YSOs are not yet certain.
They can be disk thermal instability (Lin et al. 1985; Bell \& Lin 1994; Hirose 2015), gravitational instability and disk fragmentation (Vorobyov \& Basu 2010, 2015), a combination of magneto-rotational instability and gravitational instability which also subsequently trigger thermal instability (Zhu et al. 2009), planet-disk interactions (Nayakshin \& Lodato 2012), or due to the encounters of stellar companions (Bonnell \& Bastien 1992; Pfalzner 2008). 
These mechanisms are not mutually exclusive, but rather being suitable for different size scales and physical conditions.
Resolving properties of the circumstellar disks around FU Orionis objects and EXors on various spatial scales helps diagnose the probable outburst-triggering mechanisms in individual YSOs.

Whether or not the disk gravitational instability is likely to occur may be addressed by the resolved gas and dust mass distribution in a disk. 
However, FU Orionis objects and EXors have not yet been well resolved in (sub)millimeter observations of dust spectral energy distributions (c.f., Cieza et al. 2016; Ru{\'{\i}}z-Rodr{\'{\i}}guez et al. 2017), which may provide estimates for the overall gas mass.
Previous Submillimeter Array (SMA)\footnote{The Submillimeter Array is a joint project between the Smithsonian
Astrophysical Observatory and the Academia Sinica Institute of Astronomy
and Astrophysics, and is funded by the Smithsonian Institution and the
Academia Sinica (Ho et al. 2004).} observations of four EXors suggested that the short duration and repetitive accretion outbursts may not necessarily be triggered by disk gravitational instability (Liu et al. 2016a).
The same conclusion was also given by the previous Atacama Pathfinder EXperiment (APEX) observations of CO 3-2 towards EX Lupi (K\'osp\'al et al. 2016).

The triggering mechanism(s) for the long duration outburst is/are less clear.
By performing the 0$\farcs$15 resolution Combined Array for Research in Millimeter-wave Astronomy (CARMA) observations towards the FU Orionis-like object PP\,13S*, Perez et al. (2010) suggested that surface density of the PP\,13S* disk is ten times lower than the required density to trigger disk gravitational instability.
However, they also reported that the PP\,13S* disk is optically thick inside the 48 au radius at shorter than 1.3 mm wavelengths.
Therefore, the estimates of disk mass may be uncertain. 
Dunham et al. (2012) did not detect the FU Orionis object HBC\,722 using the SMA at 1.2 mm band, and based on assumptions of optically thin and certain dust opacity and temperature deduced that HBC\,722 has a too low disk gas and dust mass to trigger gravitational instability.
However, it is not yet possible to rule out that the disk around HBC\,722 is optically thick but has a much smaller sizescale than the synthesized beam of the SMA observations, thus escaped from being detected.
On the other hand, the previous CO observations have shown that FU Orionis objects may be associated with extended gaseous disk (K\'osp\'al 2011; Hales et al. 2015). 
The high angular resolution near infrared coronagraphic polarization imaging of the previous Subaru-8.2m telescope observations further resolved spiral arm-like patterns and large-scale ($>$500 au) arcs on the disks of the four FU Orionis objects FU Ori, Z\,CMa, V1057\,Cyg, and V1735\,Cyg, which can be explained by the signatures of disk gravitational instability (Liu et al. 2016b; Dong et al. 2016).
However, the CO and infrared emission are both optically thick in the observed regions, and therefore cannot provide good constraints on disk mass distributions. 

To better understand the detailed disk property of FU Ori, we have performed high angular resolution 29-37 GHz and 8-10 GHz observations using the NRAO\footnote{The National Radio Astronomy Observatory is a facility of the National Science Foundation operated under cooperative agreement by Associated Universities, Inc.} Karl G. Jansky Very Large Array (JVLA).
In addition, we have performed 224 GHz and 272 GHz observations using the SMA.
Our observations are introduced in Section \ref{sec:obs}.
The results are presented in Section \ref{sec:result}.
We modeled the spectral energy distributions (SEDs) of the resolved sources, and derived the lower limits of the disk gas and dust mass, which are discussed in Section \ref{sec:discussion}.
In addition, we introduce our hypothesis to explain all observed features by far.
Our conclusion is given in Section \ref{sec:conclusion}.

Through out this manuscript, we assume the distance of FU Ori to be $d$$\sim$353($^{+81}_{-56}$) pc, according to the parallax measurement published in the first data release of the Gaia space telescope (Gaia Collaboration 2016).
Based on the updated distance, we have accordingly corrected the previously measured physical quantities which were based on the $d$$\sim$450 pc assumption, if it is not specifically mentioned in the related discussion.

\begin{figure}
\vspace{-0.1cm}
\hspace{-0.3cm}
\includegraphics[width=9.5cm]{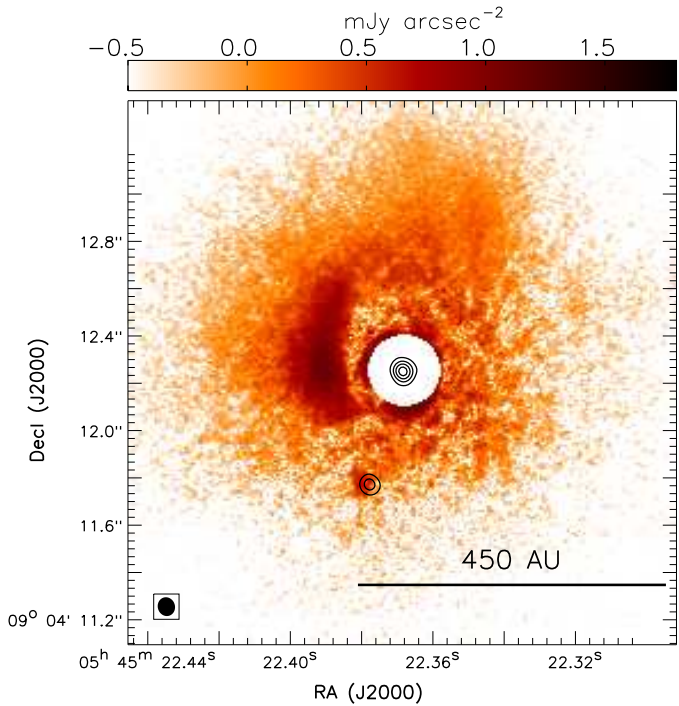}
\vspace{-0.2cm}
\caption{JVLA 33 GHz image (contour), overlaid with the Subaru-HiCIAO H-band (1.6 $\mu$m) polarized intensity image (color; see Liu et al. 2016b).
The synthesized beam of the JVLA observations is plotted in bottom left.
Contour levels are 36 $\mu$Jy\,beam$^{-1}$ (10$\sigma$) $\times$ [1, 2, 3, 4, 5].
The central 0$\farcs$3 region of the Subaru-HiCIAO image is masked due to the obscuration of the coronagraph.
We assume a distance of 353 pc, which is different from the distance quoted in Liu et al. (2016b).
}
\label{fig:hiciao}
\end{figure}

\section{Observations}\label{sec:obs}
\subsection{JVLA Observations}
\subsubsection{Ka band (29-37 GHz)}\label{subsub:ka}
We have performed JVLA Ka band (29-37 GHz) standard continuum mode observations toward FU Ori in the A array configurations on 2016 October 06 and 11 (project code: 16B-080). 
We took full RR, RL, LR, and LL correlator products. 
These observations had an overall duration of 6 hours, with $\sim$2.2 hours of integration on the target source. 
The observations on October 06 and 11 covered the parallactic angle ranges of 313$^{\circ}$-365$^{\circ}$ and 328$^{\circ}$-391$^{\circ}$, respectively.
After initial data flagging, 25 antennas were available. 
The projected baseline lengths covered by these observations are in the range of 650–36600 m ($\sim$72-4026 $k\lambda$), which yielded a $\theta_{\mbox{\tiny{maj}}}$$\times$$\theta_{\mbox{\tiny{min}}}$=0$\farcs$080$\times$0$\farcs$071 (P.A.=15$^{\circ}$) synthesized beam, and a maximum detectable angular scale of $\sim$1$\farcs$4.
We used the 3-bit sampler and configured the backend to have an 8 GHz simultaneous bandwidth coverage by 64 consecutive spectral windows, which were centered on 33 GHz sky frequency. 
The pointing and phase referencing centers for our target source is on R.A. = 05$^{\mbox{\scriptsize{h}}}$45$^{\mbox{\scriptsize{m}}}$22$^{\mbox{\scriptsize{s}}}$.357 (J2000), decl. =  +09$^{\circ}$04$'$12$\farcs$4 (J2000). 
Antenna pointing was calibrated approximately every one hour.
The complex gain calibrator J0532+0732 was observed for 20s every 80s to calibrate atmospheric and instrumental gain phase and amplitude fluctuation. 
We observed the bright quasar 3C147 for passband and absolute flux calibrations.

We manually followed the standard data calibration strategy using the Common Astronomy Software Applications (CASA; McMullin et al. 2007) package, release 4.7.0-1. 
After implementing the antenna position corrections, weather information, gain-elevation curve, and opacity model, we bootstrapped delay fitting and passband calibrations, and then performed complex gain calibration. 
We applied the absolute flux reference to our complex gain solutions, and then applied all derived solution tables to the target source. 

We generated images using the CASA task {\tt clean}. 
The image size is 6000 pixels in each dimension, and the pixel size is 0$\farcs$01. 
To maximize sensitivity, we performed multi-frequency {\tt clean} jointly for all observed spectral windows, which yielded a $\sim$33 GHz averaged frequency and a Stokes I rms noise level of $\sim$3.6 $\mu$Jy\,beam$^{-1}$.
The measured noise level is consistent with the theoretical noise level.
We also imaged every 1 GHz of bandwidth separately (rms$\sim$10 mJy\,beam$^{-1}$), to allow measuring spectral index.

\subsubsection{X band (8-10 GHz)}
We have performed JVLA X band (8-10 GHz) continuum mode observations toward FU Ori in the A array configurations on 2016 September 30 (project code: 16B-080). 
We took full RR, RL, LR, and LL correlator products. 
These observations had an overall duration of 1 hours, with $\sim$0.6 hours of integration on the target source. 
After initial data flagging, 25 antennas were available. 
We used the 8-bit sampler and configured the backend to have a 2 GHz simultaneous bandwidth coverage by 16 consecutive spectral windows, which were centered on 9 GHz sky frequency. 
The pointing center for the target source is identical to that of the Ka band observations (Section \ref{subsub:ka})
The complex gain calibrator J0532+0732 was observed for 20s every 472s to calibrate atmospheric and instrumental gain phase and amplitude fluctuation. 
We observed the bright quasar 3C147 for passband and absolute flux calibrations. 

Passband, complex gain, and absolute flux calibrations were carried out manually using CASA release 4.7.0-1. 
We applied the absolute flux reference to our complex gain solutions, and then applied all derived solution tables to the target source. 
We generated images using the CASA task clean. 
Using Briggs Robust=2 weighting yielded a $\theta_{\mbox{\tiny{maj}}}$$\times$$\theta_{\mbox{\tiny{min}}}$=0$\farcs$31$\times$0$\farcs$23 (P.A.=-41$^{\circ}$) synthesized beam.
By performing multi-frequency {\tt clean} utilizing the full bandwidth coverage, we achieved an rms noise level of 4.7 $\mu$Jy\,beam$^{-1}$, which is 11\% higher than the theoretical noise level estimated assuming no radio frequency interference (RFI).
The higher noise in our X band image is likely due to the RFI flagging.

\subsection{SMA Observations}\label{sub:sma}
\subsubsection{224 GHz}\label{subsub:smaext}
We performed the SMA 1.3 mm observations in the extended array configurations on 2013 November 09 and 22. 
The observations were carried out using a track-sharing strategy. 
The details of these SMA observations have been summarized in Table 1 of Liu et al. (2016a).
The pointing and phase-referencing centers are on the stellar position of FU Ori. 
The projected baseline lengths covered by the on-source observations are in the range of 25-165 $k\lambda$, which yielded a synthesized beam of $\theta_{\mbox{\tiny{maj}}}$$\times$$\theta_{\mbox{\tiny{min}}}$=1$\farcs$5$\times$0$\farcs$76 (P.A.=70$^{\circ}$), and a maximum detectable angular scale of $\sim$4$''$ (1400 au).
These observations used the Application-Specific Integrated Circuit (ASIC) correlator, which provided  4–8 GHz intermediate frequency (IF) coverages in the upper and the lower sidebands, with 48 spectral windows in each sideband. 
The observations tracked the rest frequency of 230.538 GHz at the spectral window 22 in the upper sideband.

\begin{table}
\caption{Millimeter fluxes of FU Ori and FU Ori S}
\label{table:flux}
\hspace{-0.0cm}
\footnotesize{
\begin{tabular}{ p{1.2cm}  p{3cm} r}\hline\hline
 & & \\[0.1pt]
Frequency$^{1}$	&  UTC	& Flux$^{2}$ \\[4pt]
(GHz)		&		& ($\mu$Jy)	\\[4pt]\hline
 & & \\[0.1pt]
32.991      & 2016Oct06/2016Oct11   &  206$\pm$4.3 (FU Ori)\\[4pt]
            &                       &  104$\pm$5.6 (FU Ori S) \\[4pt] \hline
             & & \\[0.1pt]
223.776		& 2013Nov09/2013Nov22	&  (16.7$\pm$1.6)$\times$10$^{3}$ \\[4pt] \hline
 & & \\[0.1pt]
272.412  	& 2008Dec06 			& 	(35$\pm$7)$\times$10$^{3}$ \\[4pt] \hline
 & & \\[0.1pt]
345.784$^{3}$		&	2012Dec02			&   (50.1$\pm$0.3)$\times$10$^{3}$ (FU Ori) \\[4pt]
			&						& 	(21.2$\pm$0.4)$\times$10$^{3}$ (FU Ori S) \\[4pt] \hline
\end{tabular}
}
\vspace{0.4cm}
\footnotesize{ \par $^{1}$ Central frequency of the observations. \par $^{2}$ Measured fluxes from FU Ori and FU Ori S, which are only separately listed when they can be spatially differentiated. \par $^{3}$ Taken from Hales et al. (2015).}
\end{table} 

The application of Tsys information and the absolute flux, passband, and gain calibrations were carried out using the MIR IDL software package (Qi 2003). 
After calibration, the zeroth-order fitting of continuum levels and the joint (Briggs Robust = 2) weighted imaging of all continuum data were performed using the Miriad software package (Sault et al. 1995). 
The calibrated image achieved an rms noise levels of 1.6 mJy\,beam$^{-1}$ at the averaged frequency of $\sim$224 GHz. 
A typical absolute flux calibration accuracy can be $\sim$15-20\% for SMA observations.

\subsubsection{272 GHz}\label{subsub:smacompact}
We retrieved the archival SMA 1.2 mm observations taken in the compact array configuration on 2008 December 06 (PI: Tyler Bourke).
The projected baseline lengths covered by the on-source observations are in the range of 14-72 $k\lambda$.
These observations used the ASIC correlator (Section \ref{subsub:smaext}). 
The application of Tsys information and the absolute flux, passband, and gain calibrations were carried out using the MIR IDL software package (Qi 2003). 
After calibration, the zeroth-order fitting of continuum levels and the joint imaging of all continuum data were performed using the Miriad software package to produce the 272 GHz continuum image.



\section{Results}\label{sec:result}
Figure \ref{fig:fuori} shows the SMA 224 GHz image and the JVLA 33 GHz image, which were generated utilizing the full bandwidth coverages of these observations (Section \ref{subsub:ka}, \ref{subsub:smaext}).
An overlay of the JVLA 33 GHz image with the previous Subaru High Contrast Instrument for
the Subaru Next Generation Adaptive Optics (HiCIAO) 1.6 $\mu$m polarization intensity image (Liu et al. 2016b) is presented in Figure \ref{fig:hiciao}.
The 33 GHz image resolved two compact emission sources, which are immediately around FU Ori and a projectedly nearby source FU Ori S.
They center at R.A. = 05$^{\mbox{\scriptsize{h}}}$45$^{\mbox{\scriptsize{m}}}$22$^{\mbox{\scriptsize{s}}}$.368 (J2000), decl. =  +09$^{\circ}$04$'$12$\farcs$25 (J2000) and R.A. = 05$^{\mbox{\scriptsize{h}}}$45$^{\mbox{\scriptsize{m}}}$22$^{\mbox{\scriptsize{s}}}$.377 (J2000), decl. =  +09$^{\circ}$04$'$11$\farcs$77 (J2000), respectively, which are in excellent agreement with the dust components identified by Hales et al. (2015).
FU Ori S is presently considered as a $\sim$1 $M_{\odot}$ pre-main sequence star (Wang et al. 2004).
Our SMA observations do not have high enough angular resolution to differentiate these two sources.
Table \ref{table:flux} summarizes the observed fluxes from the JVLA 29-37 GHz, and the SMA $\sim$224 GHz and $\sim$272 GHz observations.
The JVLA 8-10 GHz observations show no significant detection, which constrained the 3$\sigma$ upper limit to 14 $\mu$Jy for both sources.

Fitting two-dimensional (2D) Gaussians to the 33 GHz image yielded the deconvolved size scales (FWHM) of 36$\pm$4.5 (mas) $\times$ 30$\pm$5.1 (mas) at position angle (P.A.) 7.9$^{\circ}$$\pm$66$^{\circ}$ and 25$\pm$9.2 (mas) $\times$ 13$\pm$6.5 (mas) at P.A. 100$^{\circ}$$\pm$28$^{\circ}$, respectively.
Assuming a distance of 353 pc, the deconvolved size scales of FU Ori and FU Ori S in terms of Gaussian standard deviations are 5.4 au $\times$ 4.5 au and 3.8 au $\times$ 2.0 au.
These are two orders of magnitude smaller linear size scales than the gaseous disk traced by the Subaru-HiCIAO image (Figure \ref{fig:hiciao}), and by the previous ALMA observations of CO 3-2 (Hales et al. 2015).
Assuming that the observed 33 GHz fluxes are contributed from the deconvolved projected areas of FU Ori and FU Ori S, their averaged brightness temperature at 33 GHz are $T_{\mbox{\tiny{B, 33 GHz}}}^{\mbox{\tiny{FU Ori}}}$$\sim$210$_{-51}^{+81}$ K and $T_{\mbox{\tiny{B, 33 GHz}}}^{\mbox{\tiny{FU Ori S}}}$$\sim$360$^{+780}_{-180}$ K.

We note that the JVLA observations carried out in between August 9, 2016 and November 14, 2016 were affected by a software bug, which induced delay errors. 
The delay errors yielded displaced absolute positions of the target source in the direction of elevation, of which the magnitude can be approximated by
\[
\mbox{\small{offset}} = 5\times a\times b \,\,[\mbox{\small{milli-arcsecond}}],
\]
where $a = sec^2(z)tan(z)$, $b$ is the separation of the target source from the calibrator in the elevation direction (in degrees), and $z$ is the zenith angle.
Our Ka band observations were carried out with the elevation range of $\sim$50$^{\circ}$-65$^{\circ}$; $b$ was $\lesssim$1$^{\circ}$ at lower than 62$^{\circ}$ elevation, which gradually increased to the maximum value of $\sim$2$^{\circ}$ at $\sim$65$^{\circ}$ elevation.
The largest position displacements during our Ka band observations of FU Ori, is $\sim$0$\farcs$007.
The resulting bias in the measured source sizes is comparable, or smaller than 0$\farcs$007 given that such position displacements only smear the source geometry when the parallactic angle changes considerably (Section \ref{subsub:ka}).
The bias in source sizes can be roughly characterized by the error bars we gave, although we have to bear in mind that the errors in the measured source sizes are not symmetric but instead are negatively skewed.
Our X band observations were carried out at $\sim$45$^{\circ}$-51$^{\circ}$ elevation with $b$$\sim$1$^{\circ}$.
The delay errors did not seriously affect our X band observations due to the lower angular resolution at 8-10 GHz.

The 33 GHz emission around FU Ori is not very elongated before and after deconvolution, and there is no observed evidence for the presence of an ionized jet.
The absence of thermal radio jet was also reported by the previous, less sensitive VLA 8.5 GHz observations (Rodr\'iguez 1990).
Assuming a distance of 353 pc and a measured bolometric luminosity of 139 $L_\odot$ (Zhu et al. 2007) for FU Ori, we can use the bolometric luminosity-radio luminosity correlation of Anglada et al. (2015) to estimate that a radio flux density of 1.24 mJy is expected at 8 GHz. 
This is a factor of 90 above the 3-$\sigma$ upper limit of 0.014 mJy obtained by us. Clearly, something is strongly suppressing the ionized jet activity in FU Ori.
We also refer to the report of no enhancement of thermal radio jet emission from the recent Class 0 FU Orionis outburst of HOPS 383 (Galv\'an-Madrid et al. 2015).
Liu et al. (2014) also reported that the radio emission of Class 0/I objects may be bimodal, while the exact reason for such radio flux bimodality is not yet certain.

We note that the previous Very Large Telescope Interferometer (VLTI) Mid-Infrared Interferometric Instrument (MIDI) observations towards FU Ori have suggested a hot inner disk, which has 1.8-2.0 au and 0.94-1.2 au semimajor and semiminor axes (Quanz et al. 2006).
Limited by sensitivity and missing short spacing, the size of the mid-infrared disk observed by VLT-MIDI may be considered the lower limit of the disk size observed in the millimeter bands.
The previous detection of linearly polarized infrared scattered light from the $\sim$0$\farcs$07 resolution observations of Subaru-HiCIAO (Liu et al. 2016b) also suggested that FU Ori S is associated with an unresolved dusty disk (Figure \ref{fig:hiciao}).

\begin{figure}
\vspace{-0.6cm}
\hspace{0.0cm}
\includegraphics[width=9cm]{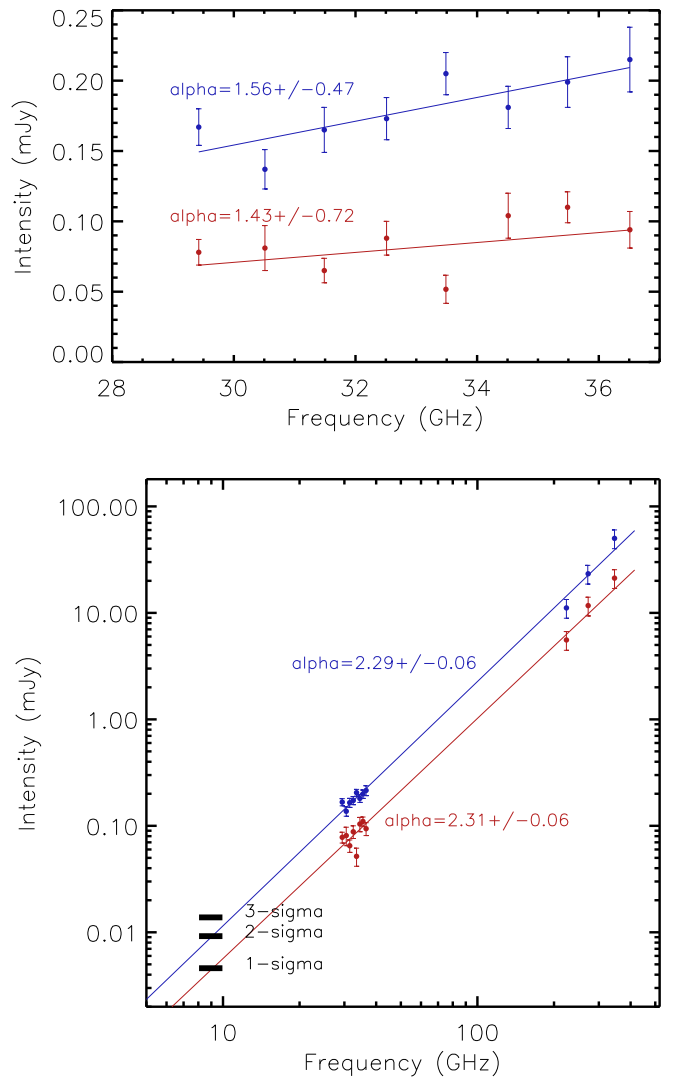}
\vspace{-1cm}
\caption{Observed fluxes from the two components, FU Ori and FU Ori S, resolved at 33 GHz (Figure \ref{fig:fuori}, right). 
Top:-- Blue and red data points are the JVLA Ka band measurements around FU Ori and FU Ori S, respectively. Solid lines are regression curves to derive spectral indices ($\alpha$).
Bottom:-- A comparison of the measured fluxes from the JVLA X band (8-10 GHz) and Ka band (29-36 GHz), the SMA 224 GHz and 272 GHz, and the ALMA 345 GHz observations. The ALMA measurements were quoted from Hales et al. (2015). We do not obtain any significant detection with the observations of JVLA at X band (8-10 GHz) and provide here the 1-3 $\sigma$ upper limits.
In the bottom panel, the plotted 224 GHz and 272 GHz data assume that the flux ratio between FU Ori and FU Ori S is identical to that at 33 GHz.  
We considered a $\sim$20\% nominal absolute flux errors for the SMA and ALMA measurements. 
We caution that the 33 GHz, 224 GHz, 272 GHz, and 346 GHz data presented here were not observed with the same $uv$ coverage. 
}
\label{fig:obsbeta}
\end{figure}

\begin{figure*}
\hspace{-1cm}
\includegraphics[width=19cm]{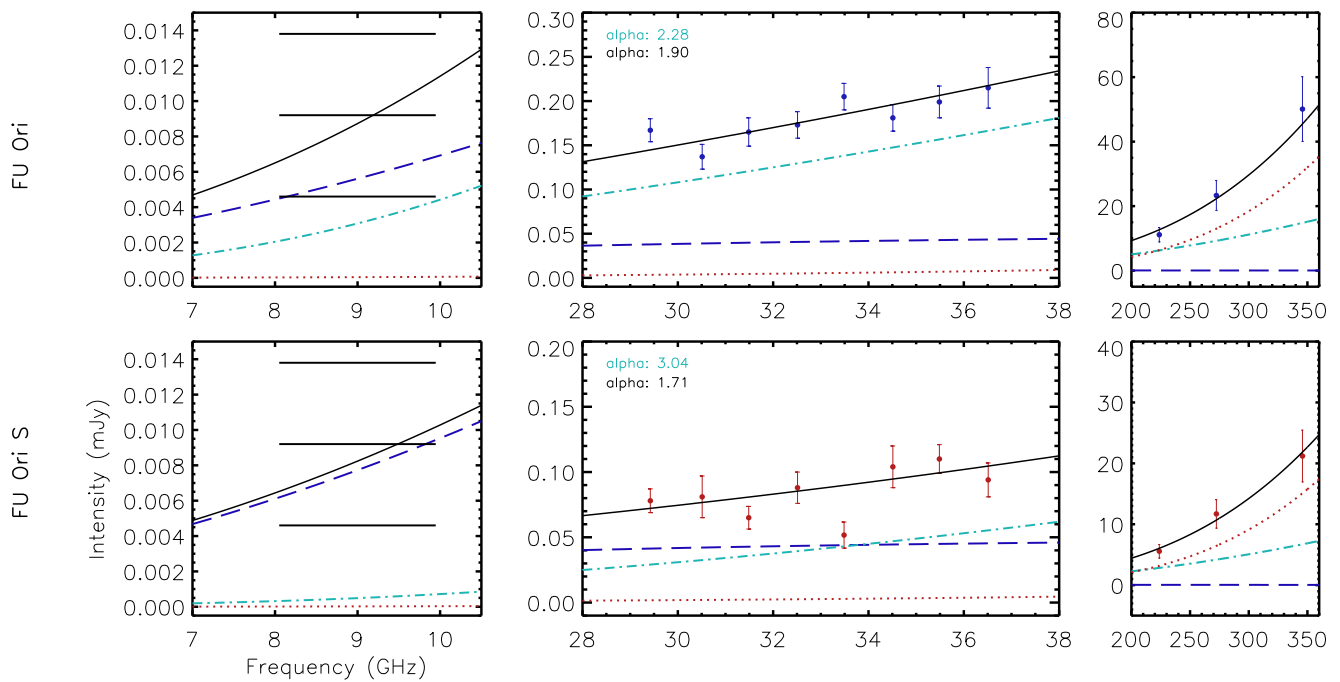}
\vspace{0.1cm}
\caption{
Fiducial SED models for FU Ori (upper) and FU Ori S (lower).
Dotted lines show the fluxes of the optical thinner and cooler dust components; dashed-dotted lines show the fluxes of the optically thicker hot dust components; dashed lines show the fluxes of the ionized components; solid curves show the integrated fluxes.
The horizontal bars in the left panels are the same with those the bottom panel of Figure \ref{fig:obsbeta}.
The plotted 224 GHz and 272 GHz data assume that the flux ratio between FU Ori and FU Ori S is identical to that at 33 GHz. 
}
\label{fig:model}
\end{figure*}


Upper panel of Figure \ref{fig:obsbeta} shows the measured 29-37 GHz SEDs of FU Ori and FU Ori S.
We performed Least Square fittings to determine their spectral indices ($\alpha$). 
We determine our fitting results and errors using a bootstrapping process.
We disturb data points at each frequency with a Gaussian random number of which the standard deviation is the 1$\sigma$ uncertainty of the measured flux.
In the end, $\alpha$ and its error were determined by the mean and standard deviations of the fittings to 10,000 independent realizations of perturbed fluxes.
The derived $\alpha$ in this frequency range for FU Ori and FU Ori S are labeled in Figure \ref{fig:obsbeta}, which are 1.56$\pm$0.47 and 1.43$\pm$0.72, respectively.
The large fitting errors of $\alpha$ are related to our small frequency coverage.
In this case, small flux errors at the highest and the lowest observed frequencies can significantly impact the fittings of $\alpha$.
The fitted $\alpha$ values are consistent with the black body emission in Rayleigh-Jeans limit to 1$\sigma$ (i.e., $\alpha$=2; for more discussion see Section \ref{sub:mass}).
Incorporating free-free emission from a little amount of ionized gas may explain the lower than 2.0 spectral indices, which however is only marginally constrained by our present frequency coverage (Section \ref{sub:mass}).
Extrapolating the overall 33 GHz flux to 224 GHz and 272 GHz assuming $\alpha$=2 can account for 86\% and 60\% of the fluxes detected by SMA at these frequencies.
Extrapolating the 33 GHz fluxes of FU Ori and FU Ori S to 346 GHz assuming $\alpha$=2 can account for 45\% and 54\% of the flux detected by ALMA (Hales et al. 2015).
We note that the typical absolute flux calibration error of ALMA at high frequency bands, and that of SMA, can be 15\%-20\%, which is not considered in the estimates of percentages here.

Jointly fitting all observed data (Figure \ref{fig:obsbeta}, bottom panel) assuming that the flux ratio between FU Ori and FU Ori S at 224 GHz and 272 GHz is identical to that at 33 GHz yielded $\alpha$=2.3$\pm$0.06 for both sources. 
The derived larger values of $\alpha$ after incorporating the SMA 272 GHz and ALMA 346 GHz measurements in the fits may be explained by the fact that these SMA and ALMA\footnote{The ALMA observations of Hales et al. (2015) covers the projected baseline lengths of 15-384 meter, which corresponds to 17-440 $k\lambda$.} observations covered shorter spacing in $uv$ sampling than the SMA 224 GHz and the JVLA observations (Section \ref{subsub:ka}, \ref{subsub:smaext}.
Therefore, the 272 GHz and 346 GHz observations may have recovered more extended fluxes (more discussion see Dunham et al. 2014).
Since these observations were separated in time by several years, we also cannot rule out the interpretation by time varying (sub)millimeter fluxes.
For an optically thick emission source, both changing temperature distribution or changing surface geometry can result in time varying fluxes.
For a few au scale circumstellar disks, the characteristic variational time scale may be as short as a few years (e.g., dynamical timescale).
On the other hand, there is not yet observational evidence for flux variability at $>$20 $\mu$m wavelengths (e.g., Green et al. 2016).
Finally, it is also possible that the JVLA, SMA, and ALMA observations have recovered all fluxes at the observed frequency bands without being affected by missing short spacing issues.
Instead, the fluxes are contributed by an optically thicker dust component, an optically thinner dust component, and potentially an ionized gas component, which dominate the fluxes at frequencies higher than $\sim$224 GHz, frequencies in between 29 GHz and 224 GHz, and frequencies lower than 10 GHz, respectively.
Such a combination naturally leads to the varying spectral index over frequency.
Example of SED models based on such hypothesis will be given in Section \ref{sec:discussion}.



\section{Discussion}\label{sec:discussion}

\subsection{Fiducial SED model}\label{sub:mass}
Around FU Ori and FU Ori S, the dominant emission mechanisms at the observed frequency bands can be thermal emission of circumstellar dust and ionized gas.
Flux of interstellar dust thermal emission as a function of frequency ($F_{\nu}$), assuming a power-law distribution of grain size, follows the modified black body formula (Hildebrand 1983):
\begin{equation}
\label{eq:dust}
F_{\nu} =B_{\nu}(T)(1-{e}^{-\tau_{\nu}})\Omega,
\end{equation}
where $\Omega$ is the observed solid angle, $\tau_{\nu}$ is the optical depth at frequency $\nu$, and $B_{\nu}(T)$$=$$(2h\nu^{3}/c^{2})$$(e^{h\nu/k_{B}T} -1)^{-1}$ is the Planck function at temperature $T$, $h$ and $k_{B}$ are Planck and Boltzmann constants. 
Dust optical depth $\tau_{\nu}$ can be further expressed by $\tau_{\nu}$$=$$\int\kappa_{\nu}\rho d\ell$, where $\kappa_{\nu}$ is the dust opacity, $\rho$ is the dust mass volume density, and $\ell$ is the optical path length.
In the case of spatially uniform $\kappa_{\nu}$, $\tau_{\nu}$=$\kappa_{\nu}$$\int\rho d\ell$=$\kappa_{\nu}$$\Sigma$, where $\Sigma$ is dust mass surface density.
The following derivations assume spatially uniform $\kappa_{\nu}$ if it is not specifically mentioned.
Opacity of interstellar dust can be approximated by $\kappa_{\nu}$$=$$\kappa_{0}$$(\nu/\nu_{0})^{\beta}$, where $\kappa_{0}$ is the dust opacity at the frequency $\nu_{0}$, and $\beta$ is the dust opacity index.
In the optically thick and Rayleigh-Jeans limit (i.e., $\tau$$\gg$1, $h\nu$$\ll$$k_{B}T$), $F_{\nu}$$\sim$$(2\nu^2/c^2)k_{B}T$$\Omega$, which has a power equal to 2 dependence on observing frequency $\nu$.
In other words, we expect to observe spectral index $\alpha$$\sim$$2$ from optically thick ($\tau$$\gg$1), high temperature black body emission sources.






Ionized gas emits free-free emission, of which the flux as a function of frequency $\nu$ is described by a similar formula with Equation \ref{eq:dust}, but with $\tau_{\nu}$ being substituted by the free-free emission optical depth $\tau_{\nu}^{ff}$.
Following Keto et al. (2003) and Mezger \& Henderson(1967), we approximate $\tau_{\nu}^{ff}$ by:
\begin{equation}
\label{eq:tauff}
\tau_{\nu}^{ff}=8.235\times10^{-2}\left(\frac{T_{e}}{\mbox{K}}\right)^{-1.35}\left(\frac{\nu}{\mbox{GHz}}\right)^{-2.1}\left(\frac{\mbox{EM}}{\mbox{pc\,cm$^{-6}$}}\right),
\end{equation}
where $T_{e}$ is the electron temperature, and EM is the emission measure defined as EM$=$$\int n_{e}^{2}d\ell$, $n_{e}$ is the electron number volume density.
Equations \ref{eq:dust} and \ref{eq:tauff} show that the spectral index of free-free emission is $\sim$2 in the optically thick limit, when the frequency dependence is dominated by Planck function.
In the optically thin limit, the spectral index is $\sim$$-$0.1.
Given a certain EM, free-free emission is optically thicker at low frequency, and is optically thinner at high frequency.

Assuming that the (sub)millimeter fluxes of FU Ori and FU Ori S are not varying with time since 2008, the simplest interpretation for their (sub)millimeter SED, and the non-detection at the 8-10 GHz band, may be optically thick dust emission, which naturally results in $\alpha$$\sim$2.
Based on this interpretation and Equation \ref{eq:dust}, we estimate the lower limit of the overall gas and dust mass to be 8.6-17$\times$10$^{-3}$ $M_{\odot}$, which is outlined in detail in Appendix \ref{appendix:mass}.
This interpretation requires attributing all observed deviations from the fitting results (which assumed constant $\alpha$ over frequency), to measurement errors.

A slightly more complex model fits better all the observations.
A composition of an ionized gas component, which emits optically thick free-free emission at $\lesssim$8-10 GHz and optically thin free-free emission at $\gtrsim$29-37 GHz, and a $\beta$$\sim$1.75 thermal dust emission source, which is optically thin at $\lesssim$8-10 GHz (e.g., $\alpha$$\sim$3.75) and is optically thick at $\gtrsim$29-37 GHz (e.g., $\alpha$$\sim$2), can result in $\alpha$$<$2.0 at 29-37 GHz and a rapidly dropping overall flux toward lower frequency such that it cannot be detected by sensitive JVLA X band observations at 8-10 GHz.
Including an additional optically thinner dust emission source, which can have a more dominant contribution to the overall flux at high frequency due to its larger spectral index, can allow the raising of $\alpha$ at $\gtrsim$224 GHz.

\begin{table*}
\caption{Parameters for SED modeling}
\label{tab:model}
\hspace{-0.1cm}
\footnotesize{
\begin{tabular}{ | l | c c c | c c c| c c c | }\hline\hline
  & & & & & & & & &   \\[0.2pt]
	&  $T_{e}$	& EM & $\Omega^{ff}$ & $T^{\mbox{\tiny{thicker}}}$ & $\Omega^{\mbox{\tiny{thicker}}}$ & $\kappa_{\mbox{\tiny{230 GHz}}}$$\Sigma^{\mbox{\tiny{thicker}}}$ & $T^{\mbox{\tiny{thinner}}}$ & $\Omega^{\mbox{\tiny{thinner}}}$ & $\kappa_{\mbox{\tiny{230 GHz}}}$$\Sigma^{\mbox{\tiny{thinner}}}$ \\[5pt]
    & (1) & (2) & (3) & (4) & (5) & (6) & (7) & (8) & (9) \\[5pt]
    & (10$^{2}$ K) & (cm$^{-6}$\,pc) & (sr) & (10$^{2}$ K) & (sr) &  & (10$^{2}$ K) & (sr) & \\[5pt]\hline
  & & & & & & & & &   \\[1pt]
FU Ori   & 160 & 6.98$\times$$10^{9}$ & 1.41$\times$$10^{-16}$ & 3.0 & 1.38$\times$$10^{-14}$ & 103 & 0.6 & 3.88$\times$$10^{-12}$ & 2.06$\times$$10^{-2}$ \\[6pt]
FU Ori S & 160 & 4.85$\times$$10^{9}$ & 1.94$\times$$10^{-16}$ & 3.6 & 5.19$\times$$10^{-15}$ & 32 & 0.6 & 1.04$\times$$10^{-12}$ & 3.87$\times$$10^{-2}$  \\[6pt]\hline
\end{tabular}
}
\vspace{0.3cm}
\footnotesize{ 
\par We use dust opacity index $\beta$=1.75 for the optically thicker and the optically thinner dust components. \par (1) Electron temperature. (2) Emission measure of the ionized gas component. (3) Solid angle of the ionized gas component. (4) Temperature of the optically thicker dust thermal emission component. (5) Solid angle of the optically thicker dust thermal emission component. 1 sr $\sim$4.25$\times$10$^{10}$ square arcsecond. (6) The dust mass surface density multiplied by the dust opacity at 230 GHz for the optically thicker dust thermal emission component, which is dimensionless. 
(7) Temperature of the optically thinner dust thermal emission component. (8) Solid angle of the optically thinner dust thermal emission component. (9) The dust mass surface density multiplied by the dust opacity at 230 GHz for the optically thinner dust thermal emission component, which is dimensionless.
 }
\end{table*}

Examples of such three-components SED models for FU Ori and FU Ori S based on this concept, are provided in Figure \ref{fig:model}.
The detailed parameters in our fiducial models are summarized in Table \ref{tab:model}.
We fix the dust opacity index to $\beta$=1.75, which is commonly observed from the dense molecular clouds and star-forming cores.
The relatively large value of $\beta$ (as compared with T Tauri disks) we choose here is to avoid detectable flux at X band.
In our models, the overall fluxes for each of FU Ori and FU Ori S at all frequencies were evaluated by simply co-adding the fluxes of a free-free emission component, an optically thicker dust thermal emission component, and an optically thinner dust component.
The derived optical depths of the free-free emission component at 8 GHz and 33 GHz, and those of the optically thicker and thinner dust thermal emission component, are summarized in Table \ref{tab:tau}.
We note that since all of the presented JVLA, SMA and ALMA observations cannot spatially resolve the sources, and also because that our observations do not provide enough of independent sampling in frequency domain, the parameters $\kappa_{\mbox{\tiny{230 GHz}}}$$\Sigma$, $T$, and $\Omega$ are degenerated. 
The values of $\beta$ for the optically thicker dust components is not certain, since our observations at 8-10 GHz only provide upper limits of fluxes. 
The non-detection of FU Ori at 8-10 GHz however prefers $\beta$$\gtrsim$1.5, such that the dust emission drops rapid enough toward lower frequency.
We note that the 1-3 $\sigma$ limits provided in the left panels of Figure \ref{fig:model} were derived assuming imaging with a full 2 GHz bandwidth.
Averaging the modeled 8-10 GHz flux of FU Ori will yield $<$2$\sigma$ significance when comparing with our detection limits.
On the other hand, extrapolating the observed flux of FU Ori at 33 GHz (Table \ref{table:flux}) to 9 GHz assuming $\alpha$=2 will yield $F_{\mbox{\tiny{9 GHz}}}^{\mbox{\tiny{FU Ori}}}$=15 $\mu$Jy, which is slightly above our 3$\sigma$ detection limit.
Therefore, we do not think the observations of FU Ori at 8-10 GHz and 29-37 GHz can be explained only with free-free emission since the spectral index of free-free emission can only be smaller than 2, although this observational constraint is yet marginal.
The fitting for FU Ori S is less constrained by our detection limit at 8-10 GHz since its observed flux at 33 GHz is lower, which allows higher fractional contribution from $\alpha$$\sim$2 emission components (e.g., relatively optically thick free-free emission or $\beta$$\rightarrow$0 dust).

For these particular SED models, the choices of $\Omega$ for the optically thicker dust components were motivated by the deconvolved size scales of FU Ori and FU Ori S at 33 GHz, and assumed that the optically thicker dust component is spatially more compact than the optically thinner dust component.
According to the previous ALMA observations at 346 GHz (Hales et al. 2015), we set the optically thinner dust components around FU Ori and FU Ori S to have sub-arcsecond angular scales. 
The degeneracy in our SED modeling can allow reducing the temperatures of the optically thinner dust components by a factor of 2 while increasing the linear size scales of them by a factor of $\sqrt{2}$, which still leaves the parameters in very reasonable ranges.
The size-temperature degeneracy of the optically thinner dust components can be resolved by future, higher angular resolution ALMA observations at $<$1 mm bands (e.g., ALMA band 6-9).
The ionized gas components are required to be spatially very compact, such that they do not contribute to detectable fluxes in the optically thick regime at 8-10 GHz.
Our present models do not uniquely explain the measured fluxes, but serve as simplified cases for qualitatively comprehending the observational results.
In the case that the emission components are optically thick, the precise evaluation of overall fluxes also need to consider more carefully the radiative transfer effects, which requires ray-tracing modeling and is beyond the possibilities of our present data due to degeneracy.
Sources with the same geometric configuration and temperature profile as FU Ori but are observed in close to  edge-on projection, may present very different (sub)millimeter SEDs, since the presumably centrally embedded free-free emission component and the optically thicker hot dust components may be partly or fully obscured.

We note that based on the infrared spectral line observational results, Zhu et al. (2007) has suggested the presence of an au scale hot disk surrounding FU Ori.
The 2D simulations of Zhu et al. (2009) also showed that in the inner few au radii, the surface temperature of the disk surrounding FU Ori can be several hundreds Kelvin (see also Vorobyov et al. 2014), which may be consistent with our derived high brightness temperatures (Section \ref{sec:result}), and the assumed dust temperatures in our SED models.
The simulations of Zhu et al. (2009) suggested that when the thermal instability is triggered, the inner $\lesssim$0.1 au scale disk can reach a gas mass density of $\sim$10$^{-8}$ g\,cm$^{-3}$, and a gas temperature of 10$^{4}$ K (see also Hirose 2015).
When dust is sublimated at such a high temperature, the disk is no more well self-shielded against the ionizing photons.
Therefore, it may be possible that the inner $\lesssim$0.1 au disk becomes  partially ionized due to stellar irradiation and viscous heating.
Dust sublimation on its own can also increase the ionization fraction because dust particles may have electrons stuck to their surface.
A line-of-sight averaged ionization fraction of $\gtrsim$10$^{-15}$, which may be plausible (e.g., Dudorov \& Khaibrakhmanov 2014), is already adequate for explaining the assumed EMs in our SED models. 
Our SED models are optically thick at 8-10 GHz, and therefore the fluxes observed in this frequency range is not sensitive to the exact ionization fraction, but are more sensitive to $T_{e}$ and $\Omega$.
However, a too high ionization fraction will make the free-free emission become optically thick also at 29-37 GHz, which may seem contradictory with the measured $<$2.0 spectral indices (Figure \ref{fig:obsbeta}).
Empirically, photo-ionized wind is too faint to be detected at the distance of FU Ori (Galv\'an-Madrid et al. 2014), although this can be uncertain.
It is also less trivial to confine the photo-ionized wind to a characteristic sub-au size scale.
The very sensitive (rms$\sim$sub-$\mu$Jy) and high angular resolution (1 mas) centimeter band observations (e.g., connecting the European VLBI Network with the Square Kilometer Array) may directly image the ionized gas component, which will help address its origin.

\subsubsection{Dust mass and the inferred gas mass}
For the model described by Table \ref{tab:model}, assuming a 100 gas-to-dust mass ratio, distance $d$=353 pc, the implied gas mass of the optically thicker and thinner dust components of FU Ori, multiplied by dust opacity at 230 GHz ($\kappa_{\mbox{\tiny{230 GHz}}}$) are (i.e., $m_{\mbox{\tiny{gas}}}$$\times$$\kappa_{\mbox{\tiny{230 GHz}}}$) 0.086 and 0.0048 $M_{\odot}$cm$^{2}$g$^{-1}$, respectively.
The values of $m_{\mbox{\tiny{gas}}}$$\times$$\kappa_{\mbox{\tiny{230 GHz}}}$ for the FU Ori S dust components are 0.0098 and 0.0024 $M_{\odot}$cm$^{2}$g$^{-1}$.
The probable values of $\kappa_{\mbox{\tiny{230 GHz}}}$ may be in between 0.1 cm$^{2}$g$^{-1}$ and 2 cm$^{2}$g$^{-1}$ (e.g., Draine 2006).
Depending on the assumption of $\kappa_{\mbox{\tiny{230 GHz}}}$, the overall gas mass can be as high as a fraction of solar mass.
If the gas mass is indeed this high, disk gravitational instability will likely be triggered (e.g., Figure 1 of Vorobyov 2013).
Considering the relatively extreme parameter space may reduce the deduced overall gas and dust mass.
For example, we may increase the temperature of the optically thick dust component of FU Ori to 1,500 K, which will reduce its solid angle and thus the dust mass by a factor of 5.
Dust starts to sublimate at higher temperature.
The conceptual design of the Next Generation Very Large Array may be ideal for providing better constraints for the disk size with a 10 times improved angular resolution (Isella et al. 2015).
Otherwise, the long baseline observations of ALMA at the 90-200 GHz (band 3, 4, 5) frequency, where the optically thicker dust component is likely brighter than the optically thinner dust component, may be considered.
An alternative way to significantly reduce the overall dust and gas mass is to reduce the gas-to-dust mass ratio, which is very difficult to be tested with observations.
We note that the summed gas mass implied by the optically thinner dust components associated with FU Ori and FU Ori S are very well consistent with the previous estimates of Hales et al. (2015)\footnote{Based on the optically thin assumption, Hales et al. (2015) derived a 1.7$\times$10$^{-4}$ $M_{\odot}$ overall dust mass.}, which were derived based on the optically thin assumption. 
However, our estimates show that the majority of the overall gas and dust mass can be contributed from the optically thicker component of FU Ori.

Deeper observations in the $<$29 GHz bands may test whether or not $\alpha$ starts increasing with the observing wavelengths, due to the larger than zero $\beta$ and the dropping optical depths of the dust emission components, as long as the centimeter band emission is not confused by the not yet detected non-thermal magnetospheric emission ($\alpha$$<$0 for most of the known cases but there may be exceptions, c.f., Liu et al. 2014). 
Otherwise, $\alpha$ remains close to 2.0 because of a significant contribution of  $\beta$$\rightarrow$0 dust, or because of the dominant contribution of optically thick free-free emission in the long wavelength bands.
The observations at $>$346 GHz will provide better constraints for the mass surface density of the optically thinner dust components.
Finally, we argue that the very detailed SED shapes instead of the sparse sampling at only a few frequencies in the millimeter and submillimeter bands, may be in general required to unambiguously derive the properties of the dusty disk and constrain dust grain growth.
This is becoming feasible thanks to the high sensitivity of ALMA and JVLA, which can make good synergy with analytic estimates or numerical simulations.

\begin{table}
\caption{Evaluated optical depths from SED modeling}
\label{tab:tau}
\hspace{1.5cm}
\footnotesize{
\begin{tabular}{ c  c c}\hline\hline
 & & \\[0.1pt]
	&  FU Ori	& FU Ori S \\[5pt]\hline
 & & \\[0.1pt]
$\tau^{ff}(\mbox{\tiny{8 GHz}})$$^{a}$  & 15 &  11   \\[5pt]
$\tau^{ff}(\mbox{\tiny{33 GHz}})$$^{b}$ & 0.79 &  0.55   \\[5pt]\hline
 & & \\[0.1pt]
$\tau^{\mbox{\tiny{thicker}}}(\mbox{\tiny{8 GHz}})$$^{c}$   & 0.29 & 0.09 \\[5pt]
$\tau^{\mbox{\tiny{thicker}}}(\mbox{\tiny{33 GHz}})$$^{d}$  & 3.5 &  1.1   \\[5pt]
$\tau^{\mbox{\tiny{thicker}}}(\mbox{\tiny{346 GHz}})$$^{e}$  & 210 &  66   \\[5pt]\hline
 & & \\[0.1pt]
$\tau^{\mbox{\tiny{thinner}}}(\mbox{\tiny{8 GHz}})$$^{f}$   & 5.8$\times$10$^{-5}$ & 1.1$\times$10$^{-4}$ \\[5pt]
$\tau^{\mbox{\tiny{thinner}}}(\mbox{\tiny{33 GHz}})$$^{g}$  & 6.9$\times$10$^{-4}$ &  1.3$\times$10$^{-3}$   \\[5pt]
$\tau^{\mbox{\tiny{thinner}}}(\mbox{\tiny{346 GHz}})$$^{h}$  & 4.2$\times$10$^{-2}$ &  7.9$\times$10$^{-2}$   \\[5pt]\hline
\end{tabular}
}
\vspace{0.4cm}
\footnotesize{ 
\par $^{a}$ Optical depth of the free-free emission component at 8 GHz.
\par $^{b}$ Optical depth of the free-free emission component at 33 GHz.
\par $^{c}$ Optical depth of the optically thicker dust thermal emission component at 8 GHz.
\par $^{d}$ Optical depth of the optically thicker dust thermal emission component at 33 GHz.
\par $^{e}$ Optical depth of the optically thicker dust thermal emission component at 346 GHz.
\par $^{f}$ Optical depth of the optically thinner dust thermal emission component at 8 GHz.
\par $^{g}$ Optical depth of the optically thinner dust thermal emission component at 33 GHz.
\par $^{h}$ Optical depth of the optically thinner dust thermal emission component at 346 GHz.
 }
\vspace{-0.5cm}
\end{table} 

\subsection{A concordant scenario to explain the observed features}
The discussion in this section assumes that our present fiducial SED models are realistic, and assuming that dust and gas are well (although not necessarily perfectly) coupled.
The implied high concentration of mass in the central few au regions, the assumed $\gtrsim$1.5 dust opacity index, the inferred little amount of ionized gas in a sub-au region from our SED models,  together with the previously resolved $\sim$400 au scales spiral arm- or arc-like features from the Subaru-HiCIAO infrared polarimetric coronagraphic imaging observations (Liu et al. 2016b), may be explained by the following concordant scenario.
First, the disk gravitational instability likely caused by the interactions of the disk with FU Ori S induced the massive inward motion of gas and dust from the several hundreds au radii to the inner few au regions.
The very high density gas and dust then piled up around the few au radii due to the development of a deadzone with negligible ionization, which yielded a significantly suppressed gas viscosity.
The short timescale of the gas accumulation in the dead zone have lead only to a modest dust grain growth in the inner disk, while in the rest of disk the dust grains remain of much smaller size.
The piled-up gas and dust eventually led to 10$^{4}$ K temperature in a sub-au scale, optically thick region, allowing dust sublimation and a weak ionization fraction and ultimately triggering thermal and magneto-rotational instabilities (Lin et al. 1985; Bell \& Lin 1994; Zhu et al. 2009; Hirose 2015).
In the theoretical frameworks of thermal+magneto-rotational instability (e.g., Zhu et al. 2009), the weakly ionized gas can instantaneously enhance gas viscosity and thereby triggers protostellar accretion outburst.
We note that FU Ori S and FU Ori are presently projectedly separated by $\sim$2.4$\times$10$^{10}$ km ($\sim$160 au), and the accretion outburst of FU Ori was since 1936.
If there is a very close encounter (e.g., impact parameter from a few to a few tens au) of FU Ori and FU Ori S when the accretion outburst was triggered, it will require these two sources to have a time-averaged relative velocity of $\sim$8 km\,s$^{-1}$ or higher over the last 80 years, which appears too fast considering that the mean stellar velocity is likely on the order of $\sim$1 km\,s$^{-1}$.
Allowing a piling-up mechanism for the inward moving gas and dust may however induce a time lag between the accretion outburst and the closest encounter, which can reduce the inferred relative velocity between FU Ori and FU Ori S.
In other words, the closest encounter between FU Ori and FU Ori S may occur at some longer time before the begin of the accretion outburst of FU Ori.
Quickly accumulating dust via a fast inward motion may also support our hypothesis of modest dust grain growth and $\beta$$\gtrsim$1.5.
We refer to Ormel et al. (2009) and references therein, for the characteristic timescale of dust grain growth as a function of density.



\section{Conclusion}\label{sec:conclusion}
We have performed $\sim$0$\farcs$07 resolution 29-37 GHz and $\sim$0$\farcs$3 resolution 8-10 GHz JVLA observations, and have performed $\sim$1$''$ resolution 224 GHz and $\sim$2$''$ resolution 272 GHz SMA observations, towards the archetype long duration accretion outburst YSO FU Ori.
We do not find any evidence for the presence of thermal radio jets. 
Our 33 GHz image resolves two compact sources around FU Ori and its companion FU Ori S, of which the deconvolved radii are only a few au.
The comparison with the previous ALMA cycle-0 observations show that the spectral indices of these two sources may not be constant over the observed frequency ranges, although it can be interpreted by the time variability of (sub)millimeter fluxes. 
For FU Ori, the spectral index is larger than 2.0 at 8-29 GHz, is very close to 2.0 at 29-224 GHz, and becomes larger than 2.0 at 272-346 GHz.
For FU Ori S, the spectral index is not well constrained at 8-29 GHz, is around 2.0 at 29-224 GHz, and is larger than 2.0 at 272-346 GHz.
Such varying spectral index may also be realized by a composition of a free-free emission component, an optically thicker dust emission component, and an optically thinner dust emission component. 
If this is indeed the case, then the disks of FU Ori and FU Ori S may be weakly ionized inside the sub-au radii, which may be better constrained by deeper observations covering a broad range of frequency (e.g., $\sim$5-50 GHz).
Finally, the dust emission at 29-224 GHz around FU Ori and FU Ori S is unlikely to be fully optically thin unless the grain growth is extreme ($\beta$$\rightarrow$0).
In the case that dust emission is optically thick at 33 GHz, we estimate the lower limit of the overall gas and dust mass to be 8.6-17$\times$10$^{-3}$ $M_{\odot}$.
The more sophisticated SED fittings show that the overall gas mass in the FU Ori and FU Ori S system can be as high as a fraction of a solar mass, which can trigger disk gravitational instability.
If the gas and dust are well coupled, our SED models will imply piling up sub-solar mass of gas within a few au (or smaller) radii.

\begin{acknowledgements}
This paper makes use of the following ALMA data: ADS/JAO.ALMA\#2011.0.00548.S. 
ALMA is a partnership of ESO (representing its member states), NSF (USA) and NINS (Japan), together with NRC (Canada) and NSC and ASIAA (Taiwan), in cooperation with the Republic of Chile. 
The Joint ALMA Observatory is operated by ESO, AUI/NRAO and NAOJ.
This work has made use of data from the European Space Agency (ESA)
mission {\it Gaia} (\url{http://www.cosmos.esa.int/gaia}), processed by
the {\it Gaia} Data Processing and Analysis Consortium (DPAC,
\url{http://www.cosmos.esa.int/web/gaia/dpac/consortium}). Funding
for the DPAC has been provided by national institutions, in particular
the institutions participating in the {\it Gaia} Multilateral Agreement.
E.I.V. acknowledges support from the Austrian Science Fund (FWF) under research grant I2549-N27.
Y.H. is currently supported by JPL/Caltech under a contract from NASA. 
M.T. is partly supported by JSPS KAKENHI Grant Number 15H02063.
R.G.-M. acknowledges support from UNAM-DGAPA-PAPIIT IA101715. 
\end{acknowledgements}

\appendix
\section{Estimates of the lower limits of dust and gas masses}\label{appendix:mass}
The observed fluxes and spectral indices at $\sim$33 GHz from FU Ori and FU Ori S may be explained by the optically thick dust emission from their surrounding circumstellar disks.
If this is indeed the case, then the lower limits of dust masses in the disks are related to the assumed values of $\kappa_{\mbox{\tiny{33 GHz}}}$.
We consider three generic assumptions of dust opacity and opacity index which are motivated by Draine et al. (2006): (1) $\kappa_{\mbox{\tiny{230 GHz}}}$$=$0.35 cm$^{2}$g$^{-1}$, $\beta$=0.5, (2) $\kappa_{\mbox{\tiny{230 GHz}}}$$=$0.67 cm$^{2}$g$^{-1}$, $\beta$=1.0, and (3) $\kappa_{\mbox{\tiny{230 GHz}}}$$=$1.29 cm$^{2}$g$^{-1}$, $\beta$=1.5.
The derived $\kappa_{\mbox{\tiny{33 GHz}}}$ in these three cases are 0.13 cm$^{2}$g$^{-1}$, 0.096 cm$^{2}$g$^{-1}$, and 0.07 cm$^{2}$g$^{-1}$.
The required $\Sigma$ value to yield optical depth $\tau_{\mbox{\tiny{33 GHz}}}$$=$1 are 7.7 g\,cm$^{-2}$, 10 g\,cm$^{-2}$, and 14 g\,cm$^{-2}$, respectively.
In these cases, the integrated dust mass within 1$\sigma$ radii of the deconvolved 2D Gaussians for FU Ori (Section \ref{sec:result}) are  16$\times$10$^{28}$ g, 22$\times$10$^{28}$, and 31$\times$10$^{28}$, respectively; the integrated dust mass in the deconvolved area of FU Ori S are 5.1$\times$10$^{28}$ g, 6.6$\times$10$^{28}$, and 9.4$\times$10$^{28}$, respectively.
Assuming an interstellar gas-to-dust mass ratio of 100, the derived lower limit of overall gas mass is in the range of 8.6-17$\times$10$^{-3}$ $M_{\odot}$.
We note that gas and dust are not necessarily well coupled. 
Therefore, the predicted gas do not need to spatially coincides with the resolved emission regions at 33 GHz (Figure \ref{fig:fuori}).
However, the extended gaseous disk may be traced by the observations of (sub-)micron size dust grains, which are better coupled with gas (e.g., Figure \ref{fig:hiciao})
We also note that if the maximum dust grain size is smaller than 100 $\mu$m, then the actual dust opacity can be one order of magnitude smaller than the values we assume here (Draine et al. 2006), which will imply one order of magnitude higher dust and gas masses than our present estimates.
In addition, the $\tau_{\mbox{\tiny{33 GHz}}}$$=$1 criterion is merely a marginally optically thick condition.
It requires $\tau_{\mbox{\tiny{33 GHz}}}$ to be larger by one order of magnitude to make $\alpha$ close to 2, unless there is significant grain growth (more in Section \ref{sec:discussion}).
If $\alpha$ is not very close to 2.0, the predicted SED will show tension with the observational results at 29-37 GHz and 224 GHz, unless we consider that our JVLA observations have frequency-dependent absolute flux calibration errors.

The deconvolved size scales of FU Ori and FU Ori S at 33 GHz can be overestimated due to the limited angular resolution, or due to some residual phase calibration errors (e.g., Perez et al. 2013).
If the observed 33 GHz fluxes are contributed by the combination of optically thick emission regions and optically thin emission regions, it is also likely that the former have more compact size scales.
If the optically thick disks actually occupy smaller projected areas than the deconvolved size scales derived in Section \ref{sec:result}, then the estimated lower limits of dust and gas masses can be reduced.
However, for FU Ori, we may not be able to reduce the estimated projected areas and whereby the dust and gas masses by more than a factor of $\sim$7, which will require the average brightness temperature to be higher than the dust sublimation temperature of $\sim$1,500 K to achieve the observed total fluxes.
The observed spectral index of $\sim$2 may also be interpreted by very efficient grain growth, such that $\beta$$\rightarrow$0.
In the cases of efficient grain growth, achieving $\alpha$$\rightarrow$2 does not require dust emission to be optically thick.
On the other hand, the dust optical depth still cannot be too low, otherwise will require dust temperature to be higher than the dust sublimation temperature to achieve the observed brightness temperature.
We are not sure whether $\beta$$\rightarrow$0 can be realistic (c.f. Draine et al. 2006), in particular, for young stellar objects which are still in an embedded phase. 

Finally, given that FU Ori has been maintaining a $\sim$10$^{-4}$ $M_{\odot}$\,yr$^{-1}$ accretion rate for $\sim$80 years (Zhu et al. 2007), if the present actual gas mass in its disk is lower than the 8.6-17$\times$10$^{-3}$ $M_{\odot}$ lower limit we derive, it will imply that the disk material had been extremely centrally concentrated, and the majority of disk material was accreted onto the host protostar FU Ori over the last century.
Our estimated lower limits of disk mass may not be sufficient for the development of disk gravitational fragmentation (Vorobyov 2013), but is probably sufficient for the development of gravitational instability, especially if an external gravitational perturbation from an intruder or companion is applied to the disk (Thies et al. 2010)

\end{document}